\documentclass[journal]{IEEEtran}

\usepackage{epsfig,amsmath}
\begin{document}
%
\title{Spin-polarized transport in II--VI magnetic resonant tunneling devices}
%
%
\author{David S\'anchez, Charles Gould, Georg Schmidt, and Laurens W. Molenkamp
\thanks{Manuscript received January 7, 2007.
        This work was supported by the Darpa SpinS, ONR, SFB 410, BMBF,
        RTN and MEC.}
\thanks{The authors are with the Physikalisches Institut (EP3),
Universit\"{a}t W\"{u}rzburg, Am Hubland, D-97074 W\"{u}rzburg, Germany.
D. S\'anchez is also with the Departament de F\'{\i}sica,
Universitat de les Illes Balears, E-07122 Palma de Mallorca, Spain.}
\thanks{Copyright \copyright 2007 IEEE. Reprinted from
{\it IEEE Trans. Electron Devices,~Vol.~54, No.~5, pp. 984-990, May~2007}.
This material is posted here with permission of the IEEE. Such permission
of the IEEE does not in any way imply IEEE endorsement of any Cornell
University Library's products or services.  Internal or personal use of
this material is permitted.  However, permission to reprint/republish this
material for advertising or promotional purposes or for creating new
collective works for resale or redistribution must be obtained from the
IEEE by writing to pubs-permissions@ieee.org.
By choosing to view this document, you agree to all provisions of the
copyright laws protecting it.}}

\markboth{IEEE Transactions on Electron Devices,~Vol.~54, No.~5,~May~2007}{D. S\'anchez \MakeLowercase{\textit{et al.}}: Spin-polarized transport in magnetic resonant tunneling devices}


\maketitle

\begin{abstract}
We investigate electronic transport through II--VI semiconductor resonant
tunneling structures containing diluted magnetic impurities.
Due to the exchange interaction
between the conduction electrons and the impurities, there arises a giant
Zeeman splitting in the presence of a moderately low magnetic field.
As a consequence, when the quantum well is magnetically doped
the current--voltage characteristics shows two peaks
corresponding to transport for each spin channel.
This behavior is experimentally observed and
can be reproduced with a simple tunneling model.
The model thus allows to analyze other configurations.
First, we further increase the magnetic field,
which leads to a spin polarization
of the electronic current injected from the leads, thus giving rise to a
relative change in the current amplitude.
We demonstrate that 
the spin polarization in the emitter can be determined from such a change.
Furthermore, in the case of a magnetically doped injector 
our model shows a large increase in peak amplitude
and a shift of the resonance to higher voltages as
the external field increases. We find that this effect
arises from a combination of giant Zeeman splitting, 3-D incident
distribution and broad resonance linewidth.
\end{abstract}


\IEEEpeerreviewmaketitle

\section{Introduction}

\PARstart{P}{resent} information processing
technology is mostly based on the manipulation
of the electron charge by means of external electric fields.
In the rapidly flourishing field of \emph{spintronics}
or \emph{magnetoelectronics} the chief role is played by the electron {\em spin}.
Spintronics has already provided us with commercial devices, e.g.,
read-out heads for computer hard disks. These gadgets make use
of the giant magnetoresistance effect in metals.
In semiconductor materials, the spin orientation of conduction electrons
survives much longer time, which makes semiconductor-based spintronic devices,
to a great extent, attractive for quantum computing where the electron spin
would play the role of a qubit of information~\cite{spin}.
\begin{figure}
\centerline{
\epsfig{file=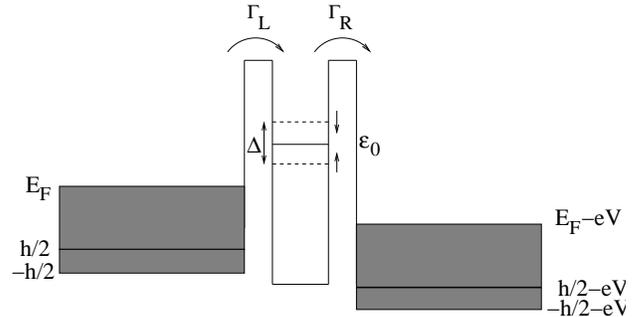,angle=0,width=0.45\textwidth,clip}}
\caption{Schematic representation of the energy profile of a
double barrier tunnel diode with a single energy level in the
quantum well, $\varepsilon_0$, coupled to two electron reservoirs
with Fermi energy $E_F$. The quantum level and the reservoir band bottom
may be spin split due to an external magnetic field.}
\label{figwell}
\end{figure}

A basic requirement for the applicability of spintronic devices
is their all-electrical ability of generating and detecting spin-polarized currents.
The most promising prospect devices use alloys of
nonmagnetic semiconductors with magnetic
transition metals (mainly, Mn) \cite{dmsreviews}. Thus, thin layers of diluted magnetic
semiconductor (DMS) compounds can be stacked to form
quantum confined geometries in which
new spin-dependent transport effects are likely to occur~\cite{ohn98}.
Recent progress has been able to reduce Mn-related scattering
and, at the same time, maximize spin effects.
For example, large band electron spin splittings
at moderately low magnetic fields can take place in
II--VI compounds (e.g., CdTe, ZnS or ZnSe) with Mn substituted
on the group II site~\cite{aws99}.
Since the valence of the cations Cd or Zn equals that of Mn,
additional n-doping is required to generate electric carriers.
The possibility of n-doping is a great advantage of II--VI-based devices
over their III-V counterparts (typically, (Ga,Mn)As) which are always p-type,
since this difference allows to significantly decrease the amount of spin-orbit
coupling experienced by the carriers
and, as a result, spin dephasing and relaxation are reduced.
Another advantage is related to the spin injection efficiency \cite{fie99,ohn99}.
The price one has to pay is that n-doped II--VI DMS's
do not show ferromagnetism. Nevertheless,
II-VI DMS structures provide an excellent framework for fundamental
studies on the dynamical interplay between spin splitting subbands,
electron transport, spin decoherence and magnetically controlled
quantum confinement.

The type of DMS structures with which we are here concerned are resonant
tunneling diodes (RTD's).
These systems have recently received attention \cite{slo03,hay00,bru98,bel05,gan06,ert06}.
A RTD is a widely studied quantum semiconductor
device which is the analog of the optical Fabry-Perot resonator \cite{cha74}.
A double barrier confines the electrons
in a quantum well between two electron reservoirs.
(A simplified sketch of the conduction band
profile of a RTD sample is shown in Fig. \ref{figwell}).
Resonant tunneling arises when the incident stream of electrons matches
its energy with the quantum level formed in the well,
thereby working as an energy filter to a good extent.
A higher bias lowers the resonance energy relative to the energy of
the injected electrons and the current--voltage ($I$--$V$) thus exhibits
a negative differential conductance region.
This effect can be used for high speed switching applications in
electronic circuitry at room temperature \cite{bro87,ozb93}.

When the quantum well is doped with Mn ions, the resulting DMS is paramagnetic.
In the absence of magnetic fields, the well subbands are spin degenerate.
When a small magnetic field is applied, a {\em giant} Zeeman splitting
(of the order of a few meV)
arises due to the exchange interaction between the localized magnetic moments
(the Mn$^{++}$ ions) and the conduction band electrons \cite{fur88}.
As a consequence, the bottom of the
well energy subbands splits and two conduction channels (one per spin)
are able to participate in the transport, see Fig. \ref{figwell}.
Using bias voltage to select a
particular transmission resonance leads to the generation of a
current peak with predominantly spins up or down.
Reference~\cite{slo03} showed that each resonance appears
as a distinct peak in the $I$--$V$ characteristic and that
further increase of the magnetic field leads to a larger
peak splitting. The separation follows a modified Brillouin
function for paramagnetic systems \cite{slo03},
\begin{equation}\label{eq_delta}
\Delta=N_{\rm Mn} J_{sd} S_0 B_S
\left(\frac{S g \mu_B B}{k_B(T+T_{\rm eff})}\right)\,,
\end{equation}
where $N_{\rm Mn}$ is the concentration of $S=5/2$ Mn spins,
$J_{sd}$ is related to the exchange integral (see below),
$B_S$ is the Brillouin function with $g$ the Land\'e factor,
$B$ is the magnetic field, $T$ is the temperature,
and $S_0$ and $T_{\rm eff}$
are the Mn effective spin and temperature, respectively.
Each peak of the splitting is dominated by tunneling via a
given spin channel. As a consequence, the RTD works as a {\em spin filter}.
At large magnetic fields, the splitting reaches saturation
but the Zeeman spin splitting in the contacts, which is negligible
at low magnetic fields, starts to play a role.
The well is progressively injected with an increasing carrier
density with a given spin direction and, as a result,
the peak {\em amplitude} is modified.
In this operation regime, we predict that the DMS RTD can be regarded as
a {\em detector} of spin polarized currents.

When the quantum well is a normal semiconductor and
the injector is doped with Mn ions, a different scenario results.
The giant Zeeman splitting develops now in the injector,
which modifies the energy distribution of the injected current.
The calculated $I$--$V$ consists of a single peak with an amplitude
which enhances largely with increasing magnetic field.
Moreover, we observe a shift of the resonance to higher voltages.

Below, we discuss in detail the two predictions sketched above.
We present a theoretical model which shows that the effect
causing the peak current enhancement in the DMS
injector case is due to a combination of three-dimensional incident
distribution, giant Zeeman spin splitting and broad resonance
linewidth. In the case of a DMS quantum well, 
we demonstrate that the spin polarization of the emitter can be
determined from the relative change of the peaks forming
the splitting.

\section{Experimental sample}

\begin{figure}
\epsfig{file=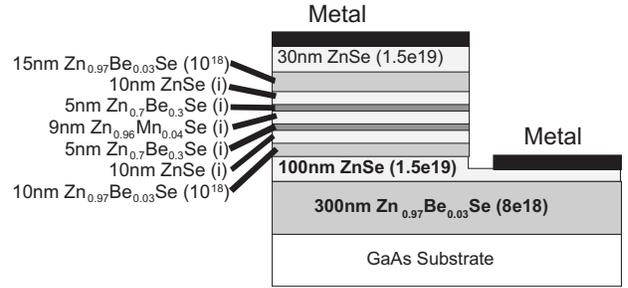,angle=0,width=0.45\textwidth,clip}
\caption{(Color online) Details of the layer stack used in the experiments.}
\label{figsample}
\end{figure}

Figure \ref{figsample} shows a pictorial representation
of the sample used in the experiment.
The tunneling region consists of a 9 nm thick Zn$_{.96}$Mn$_{.04}$Se
quantum well
sandwiched between two 5 nm thick Zn$_{.7}$Be$_{.3}$Se barriers.
The well is coupled to two 10 nm thick ZnSe layers.
The remaining
layers are needed to ensure a proper doping profile and to allow for
the fitting of high quality ohmic contacts.
The contact resistance of the devices is kept to a minimum by using
an in-situ Al (10 nm)/Ti (10 nm)/Au (30 nm) top contact, while the
ex-situ bottom contact is fabricated by etching down to the very
highly doped ZnSe layer, and using large area (500$^{2} \mu
$m$^{2})$ Ti-Au contact pad.
The injector and the collector are lattice matched to the GaAs
substrate. The barrier layers are obviously not lattice
matched but are sufficiently thin to grow fully strained.
The sample is patterned in 100
$\mu$m$^{2}$ pillars using standard optical lithography
with positive photoresist followed by metal evaporation
and lift-off.

Measurements are performed in a $^4$He bath cryostat equipped with a high field
superconducting magnet using standard DC transport techniques. Care
was taken to construct a circuit with a low (40 $\Omega$) resistor in
parallel to the diode to prevent the diode from going into
oscillations in the negative differential conductance region.
\cite{Leadb}. Current measurements consist of measuring the voltage drop
across a relatively small 30 Ohm series resistor.

\section{Experimental observations}

\begin{figure}
\epsfig{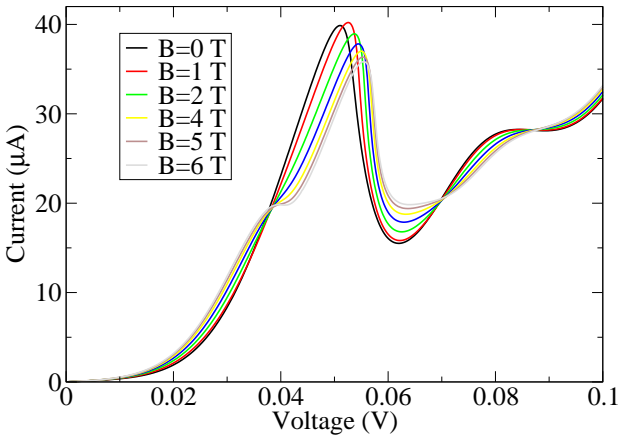}
\caption{(Color online) Experimental magnetic field dependence
of the $I$--$V$ curves
of a RTD with magnetic quantum well. The field is varied from
0 to 6~T. Temperature is 1.3~K.}
\label{figrtd}
\end{figure}

$I$--$V$ characteristics for the DMS RTD and different strengths of applied magnetic fields
are shown in Fig. \ref{figrtd}. At zero field,
the sample exhibits typical RTD behavior,
showing a strong resonance peak at 51 mV with a peak-to-valley
ratio just over 1. The additional resonance visible
at 84 mV in the $B=0$ curve is the well
known LO phonon replica \cite{Leadb} which is separated from the
direct resonance by the energy of the LO phonon of the well material,
and can be used to calibrate the voltage scale to the energy of the levels
in the quantum well. (The voltage axis shown in the experimental results
is accordingly rescaled with a lever arm of approximately $\mathcal{L}=2.1$).
A resonance associated with the second well level
occurs at higher bias (not shown in the figure). Because of the intrinsic
asymmetry of the device due to the presence of the GaAs substrate,
the resonances for the negative bias direction are not as pronounced.
We confirmed the absence of charging effects in the device
by verifying that $I$--$V$ curves for different sweep direction
were identical \cite{Martin}.

As the magnetic field (parallel to the current) is applied,
the resonance splits into two parts and the splitting grows
as $B$ increases. The growing saturates at large fields \cite{slo03}
which is consistent with the Brillouin like behavior expected from the giant
Zeeman splitting of the injector [Eq.~(\ref{eq_delta})].
The results for $B$ perpendicular to the current
are similar but they show a broader resonance as $B$ increases.
This effect is well known in GaAs RTD's and can be understood
from conservation during tunneling of the momentum parallel to the
interfaces \cite{ben88}. Let $z$ be the growth direction
and the in-plane $B$ point along $x$. Then,
the Lorentz force acting on an electron traversing
the barrier of width $d$ provides an extra electron
momentum $k_y=eBd/\hbar$. The corresponding energy shift,
$\delta E=e^2B^2d^2/2m$, is compensated at resonance with
a larger voltage $\Delta V=\delta E/e$.
Furthermore, the resonance peak is broadened for increasing $B$
since electrons with different $k_y$ see different effective
potentials and the redistribution of electron energies
cause the additional broadening \cite{ben90}.

Moreover, we obtained the temperature dependence of the
$I$--$V$ characteristics \cite{slo03} and found that
an increase in temperature has a similar effect on the curves
as a lowering of $B$. This behavior agrees
with the Brillouin function describing the magnetization.

\section{Theoretical model}

The exchange interaction that couples $s$ conduction band electrons
and $d$ Mn local moments is ferromagnetic and favors parallel alignment of
the local moment $S$ and band electron spins $s$. This is well
described by a phenomenological exchange model:
\begin{equation}
{\cal H}_{\rm int}^{sd}=\tilde{J}_{sd} \sum_{I} \vec{S}_I \cdot \vec{s}(\vec{r}_I) \,,
\label{eq-hjsS}  
\end{equation}
where the sum is extended over the positions $I$ of the magnetic impurities and
$\tilde{J}_{sd}$ is the exchange integral (which we take as a constant).
We can simplify this Hamiltonian combining 
a virtual crystal approximation with
a mean-field theory \cite{fur88,die00}.
Both of them are good approaches since, first,
the lattice parameters and the band Hamiltonian parameters
of a II-VI DMS heterostructure are assumed to change smoothly
as Mn$^{++}$ spins are introduced in the system
and, second, a $S=5/2$ quantum spin is assumed to be added to the
low energy degrees of freedom for each Mn spin.
The virtual crystal approximation consists of
placing one magnetic ion at each lattice site,
reducing the exchange constant by a factor $x$.
This is a good approximation since in semiconductors
the electron wavelength is large and the electrons
interact simultaneously with a large number of
impurities. Equation~(\ref{eq-hjsS}) is thus transformed into
\begin{equation}
{\cal H}_{\rm int}^{sd} = x \tilde{J}_{sd} \sum_{i} \vec{S}_i \cdot \vec{s}(\vec{r}_i) \,,
\label{eq-hjsS2}  
\end{equation}
where now the sum is extended over the whole lattice.
In the mean-field approach, each Mn$^{++}$ moment is replaced by
its statistically averaged projection along the
direction of the external field (say, $z$):
\begin{equation}
{\cal H}_{\rm int}^{sd} = J_{sd} \langle S_z \rangle \sum_{i} s_z(\vec{r}_i) \,,
\label{eq-hjsS3}  
\end{equation}
where $J_{sd}\equiv x \tilde{J}_{sd}$.
$\langle S_z \rangle$ is described by the response of
a paramagnetic system of noninteracting spins under the action of a magnetic field $B$:
\begin{equation}
\langle S_z \rangle = N_{\rm Mn} S_0
B_{5/2} \left( \frac{g \mu_B B S}{k_B (T+T_{\rm eff})} \right) \,.
\label{eq-sznmn}  
\end{equation}
Therefore, when the mean-field and virtual crystal approximations are employed,
the effect of this coupling is to make the subband energies spin-dependent
according to Eq.~(\ref{eq_delta})
in those regions that contain Mn ions. 
Typical parameter values for the system under consideration
are $\tilde{J}_{sd}=0.26$~eV, $g=1.1$, $S_0=1.13$ and $T_{\rm eff}=2.24$~K
for $x=8$\% ($S_0=1.64$ and $T_{\rm eff}=1.44$~K for $x=4$\%).

Because spin effects are independent of the $B$ direction,
orbital effects are not taken into account.
We ignore spin-flip processes, so that the current density
along the $z$-direction, $J=J_\uparrow+J_\downarrow$,
are carried by the two spin subsystems {\em in parallel}.
We model the transport through the magnetic
RTD considering resonant tunneling
through a double barrier system. In the transmission formalism,
the spin-dependent current density traversing the quantum well reads,
\begin{eqnarray}\label{eq_cur}
J_\sigma&=&\frac{e\nu}{h}
\int_{eV-s h/2}^\infty dE_z dE_\bot \, T_\sigma(E_z,\varepsilon_0,V) \\ \nonumber
&\times& [f_L(E_z+E_\bot)-f_R(E_z+E_\bot)]\,,
\end{eqnarray}
with $V$ the bias voltage applied to the structure,
$\nu=m/2\pi\hbar^2$ and $s=+$ ($-$) for $\sigma=\uparrow$ ($\downarrow$).
The Fermi functions $f_L$ and $f_R$ describe the distribution
of electrons with total energy $E_z+E_\bot$ in the left and right leads
with electrochemical potentials $\mu_L=E_F+eV$ and $\mu_R=E_F$, respectively
($E_F$ is the Fermi energy).
Band-edge effects are incorporated in the lower limit of the integral.

In Eq.~(\ref{eq_cur}), the transmission $T$ conserves the momentum
parallel to the interfaces and depends, quite generally, on $E_z$, $V$
and the subband bottom energy of the quantum well which
includes the spin splitting:
$\varepsilon_0\to\varepsilon_0^\sigma\equiv\varepsilon_0-s\Delta/2$.
Close to resonance, it is a good approach \cite{but88} to take a Lorentzian shape,
\begin{equation}
T_\sigma=\frac{\Gamma_\sigma^L\Gamma_\sigma^R}
{(E_z-\varepsilon_0^\sigma)^2+\Gamma^2/4} \,,
\end{equation}
where $\Gamma_\sigma^L$ ($\Gamma_\sigma^R)$ is the partial decay width
due to coupling to lead L (R) and $\Gamma=\Gamma_L+\Gamma_R$
the total broadening per spin.
We hereinafter consider symmetric barriers
for simplicity but it is important in the
strongly nonlinear regime (i.e., around the current peak)
to take into account the energy (and voltage) dependence of the tunneling rates.
For a rectangular barrier, it reads
$\Gamma_\sigma(E_z)=\Gamma_0 \sqrt{E_z-eV+\sigma h/2}\sqrt{E_z}/E_F$ \cite{bla99}.
The barrier height is given by the conduction band offset between
ZnSe and Zn$_{0.7}$Be$_{0.3}S$e, which is approximately $0.6$~eV.

To simplify Eq.~(\ref{eq_cur}) we consider an infinitely narrow resonance
($\delta$-resonance),
$T(E_z)=2\pi (\Gamma_L\Gamma_R/\Gamma)\delta(E_z-\varepsilon_0)$.
In this limit, we find
$J_\sigma(V)=e \nu\Gamma_L\Gamma_R
(E_F+eV-E_0+\sigma\Delta/2)/\hbar\Gamma$ for
${\rm max}(E_F,E_0-E_F-\sigma\Delta/2)<eV<E_0-\sigma\Delta/2$.
This simple expression predicts a splitting of the $I$--$V$
curve as observed in the experiment.
The maximum current for each spin channel is
$J_\sigma^{\rm max}=(e\nu /\hbar)
(\Gamma_L\Gamma_R/\Gamma) E_F$. The total current peak
is given by $I_p=2e \nu E_F (1-\Delta/2E_F)/\hbar\Gamma$.
which decreases with increasing $\Delta$, as expected.
\begin{figure}
\centerline{
\epsfig{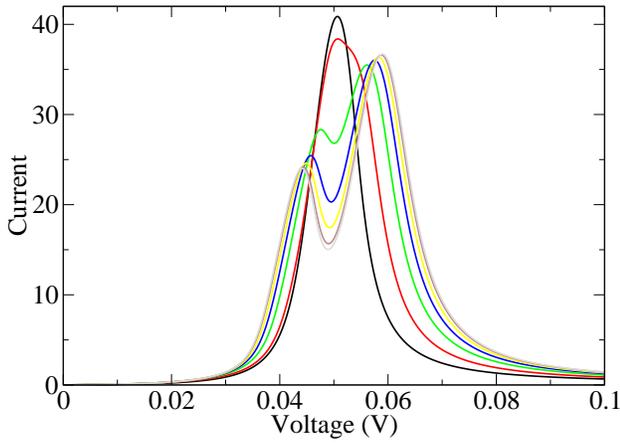}}
\caption{(Color online) Theoretical $I$--$V$ curves
at 4~K for a RTD with magnetic quantum well. The field is varied from
0 to 6~T as in Fig. \ref{figrtd}.
Parameters are $E_F=10$~meV, $\varepsilon_0=54$~meV, $\Gamma_0=2$~meV.
}
\label{figrtd_th}
\end{figure}

Numerical simulations of the full model (see Fig.~\ref{figrtd_th})
result in good agreement with the data shown before. The splitting
appears more clearly in the simulations because the nonresonant contribution
to the experimental current (which increases exponentially with $V$)
is not taken into account. In addition, the model ignores spin
relaxation effects, which tend to equilibrate
the spin subsystems and which may become important
in particular situations. Nevertheless, the simplicity of the model
suggests to explore new configurations.

\subsection{RTD with magnetic injector}
\begin{figure}
\centerline{
\epsfig{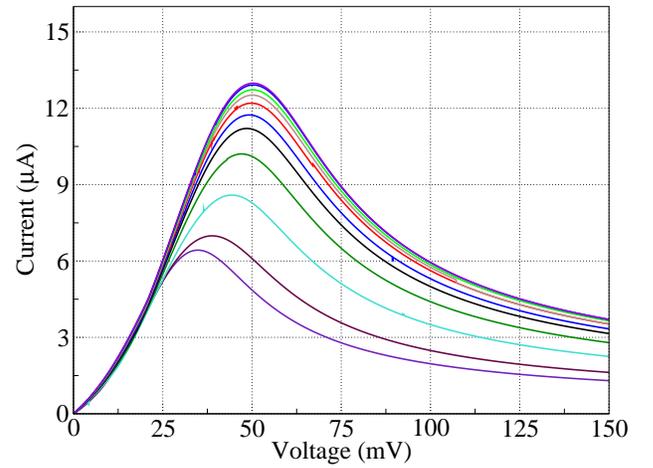}}
\caption{(Color online) Theoretical $I$--$V$ curves
at 4~K for a RTD with a spin-polarized injector increasing
the spin splitting from $h=0$ to $h=2 E_F$ in
steps of $h=0.2 E_F$ (from bottom to top).
Parameters are: $E_F=10$~meV, $\varepsilon_0=21$~meV and
$\Gamma_0=15$~meV.
}
\label{figminj_th}
\end{figure}

Let us first focus on a RTD with magnetic injector for which $\Delta=0$
in the quantum well. We denote the giant Zeeman splitting in the injector
with $h$ to distinguish it from the splitting in the well.
For large level broadening we expect that
the $I$--$V$ becomes dominated by $\Gamma$ either because
$E_F$ and temperature are quite small
or due to strong interface roughness or disorder.
Then, there should be no splitting in the RTD characteristics related
to spin-dependent transport. However, we observe an interesting effect
in the characteristics, see Fig.~\ref{figminj_th},
where the giant Zeeman splitting in the injector is varied from $h=0$ to $h=2E_F$.
Fist, the peak shifts to higher voltages and
strongly \textit{increases} in amplitude. The strongest dependence of
the peak amplitude and position
on the magnetic field is observed at low fields,
with the behavior saturating at high fields.
The behavior is therefore
consistent with the Brillouin like behavior expected from the giant
Zeeman splitting of the injector [Eq.~(\ref{eq_delta})].

The slight shift of the peak to higher bias with increasing magnetic field
is explained as follows. The maximum of the resonance occurs when the bottom
of the conduction band is brought into alignment with the well level.
As the field increases, the peak becomes more and more dominated by the
majority spin conduction band. Since the bottom of this band is moving
to lower energies, a higher bias will be required to bring this band in
alignment with the well level, producing a shift of the resonance towards
higher bias. In fact, we can expand Eq.~(\ref{eq_cur}) at zero temperature
in powers of $1/\Gamma$. We find for small spin splittings
that the resonance peaks at
\begin{equation}\label{eq_vres}
eV_{\rm res}=\varepsilon_0-\frac{E_F}{3}+\frac{h^2}{3 E_F}\,.
\end{equation}
This expression clearly shows the
shift of $V_{\rm res}$ with increasing $h$.

The strong increase in the current amplitude shown in Fig. \ref{figminj_th}
is related to the change of the conduction band energy
with magnetic field $B$ but its manifestation is somewhat subtle.
Qualitatively, this result can be viewed as resulting from the change
in the position of the conduction band in the injector which is required
to maintain the alignment of the Fermi level throughout the device while
allowing for the magnetic field induced redistribution of spin populations.
This pushes down the energy of the bottom of the majority spin conduction
band, which thus requires a higher bias to be brought into alignment
with the well level, and consequently moves the maximum of the resonance
to higher bias. Additionally, the redistribution of the spin population
increases the size of the Fermi sphere of the majority band, allowing for
more states at the Fermi energy to participate in the tunneling, and thus
increasing the tunneling current. One might naively expect that a
spin splitting of the conduction band leaves the DMS injector electron
density constant (keeping the Fermi energy fixed).
This is indeed correct for a two-dimensional injector~\cite{san02}
since the spin-dependent density at zero temperature,
$n_\sigma=(m/2\pi\hbar^2) (E_F-\varepsilon_0^\sigma)$,
is independent of energy because the density of states is constant.
However, for a three-dimensional system
$n_\sigma$ depends {\em nonlinearly} on $h$. At $k_B T=0$,
\begin{equation}
n_\sigma=\int_{-sh/2}^{E_F} \nu_\sigma (E) \,,
\end{equation}
where $D_\sigma(E)=(m/2\pi^2\hbar^2)\sqrt{2m(E+sh/2)/\hbar^2}$.
As a consequence,
the total electron density $n=n_\uparrow+n_\downarrow$,
\begin{equation}
n(h)\propto\left(1+\frac{h}{2E_F}\right)^{3/2}
+\left(1-\frac{h}{2E_F}\right)^{3/2}
\end{equation}
is an {\em increasing} function of $h$
provided $E_F$ remains fixed.
In the fully polarized case, $h=2E_F$,
the increase can be as large as 44\%.
Since the current peak is mainly determined by $n$,
it follows that the peak amplitude increases with $B$.
Therefore, the current increase effect has a {\em geometrical} origin.
Inserting Eq.~(\ref{eq_vres}) in Eq.~(\ref{eq_cur})
we find to leading order in $1/\Gamma$ that the current peak
$J_p=(em/2\pi^2\hbar^3) E_F^2 (1+h^2/4E_F^2)$
is indeed an increasing function of the magnetic field.
As a result, the RTD with magnetic injector may serve
as a generator of high peak currents by simply changing
the applied magnetic field.

\subsection{RTD with magnetic well at high fields}
We now turn back to the case of a RTD where the quantum well
is a DMS. At high fields,
while the giant Zeeman splitting
of the DMS has already reached saturation and remains
therefore unaffected by further magnetic field increases,
the normal Zeeman splitting in the nonmagnetic injector
is of course still growing linearly
with magnetic field:
\begin{equation}\label{eq_h}
h=g\mu_B B \,,
\end{equation}
For low $B$ the resulting splitting
is negligible in the RTD transport properties since it is of the order
of the thermal broadening $k_B T$. On the other hand, one usually has
$h\ll \Delta$. This situation was studied above.
Nevertheless, for high magnetic fields, $h$ can attain a
considerable fraction of the Fermi energy. E.g., at $B=14$~T, $h$ is around 1~meV
and can not be neglected for typical Fermi energies ($E_F\sim 1-10$~meV).
Therefore, as $B$ continues to increase,
a spin polarization develops in the carrier
population of the injector, and the diode
begins to be fed with spin polarized electrons.
The resonance corresponding to the lower energy
subband in the quantum well thus becomes fed
with a stronger current, and thereby increases in amplitude,
whereas the high energy spin
level in the quantum well sees its amplitude diminish as the population of injected
carriers decreases. The quantum well structure is
thus operating as a {\em spin polarized
detector} with the respective amplitude
of the spin split peaks serving as a measure of
the polarization of injector current.

In the $\delta$-resonance limit, the maximum current per spin reads,
\begin{equation}\label{eq_ism}
J_\sigma^{\rm max}=\frac{e\nu}{\hbar}\frac{\Gamma_L\Gamma_R}{\Gamma}
(E_F+s h/2)\,,
\end{equation}
which increases (decreases) for spin-up (-down) electrons.
Equation~(\ref{eq_ism}) is valid for resonant energies sufficiently
above the Fermi energy ($\varepsilon_0>E_F+\Delta/2$).
We define the relative current amplitude change when the spin splitting in
the leads is varied,
\begin{equation}\label{eq_xi}
\xi_\sigma (h)= \frac{J_\sigma^{\rm max}(h)-J_\sigma^{\rm max}(0)}
{J_\sigma^{\rm max}(0)}\,.
\end{equation}
Using Eq.~(\ref{eq_ism}) we find a short expression for the amplitude
change, $\xi_\sigma (h)= s h/2E_F$.
A particularly simple relation between the spin polarization in the injector
and $\xi$ follows from this discussion.  For free conduction
electrons we can make the following approximation for the
spin polarization: $N_\uparrow -N_\downarrow = \int_{-h/2}^{E_F} D(E) dE -
\int_{h/2}^{E_F} D(E) dE \approx D(E_F) h$,
where $D(E)$ is the density of states in the injector.
Thus, defining the polarization
$p=(N_\uparrow -N_\downarrow)/(N_\uparrow +N_\downarrow)$
and using Eqs.~(\ref{eq_ism}) and~(\ref{eq_xi}) we find $p=\xi_\uparrow$.
This demonstrates that current amplitude changes measure
the spin polarization in the injector.
\begin{figure}
\centerline{
\epsfig{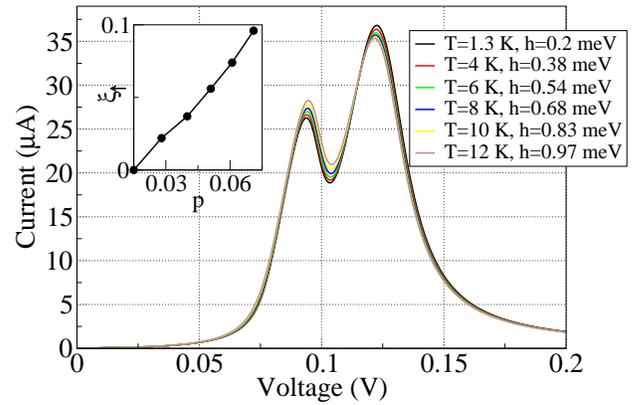}}
\caption{(Color online) Theoretical $I$--$V$ curves for the RTD
with magnetic quantum well at high fields.
The parameters are set as in Fig. \ref{figrtd_th}.
Inset: Relative change in the current amplitude of the majority spin peak
as a function of the injector spin polarization.}
\label{figmqw_th}
\end{figure}

We now numerically simulate RTD $I$--$V$ characteristics
using Eq.~(\ref{eq_cur}). 
We present the results in Fig.~\ref{figmqw_th}.
In order to better distinguish the effects of the injector
and the quantum well we vary
both the temperature and the external magnetic field between curves.
For each curve, a pair of $k_B T$ and $B$
parameters are selected such that the giant Zeeman splitting in the quantum well
is maintained at a constant value $\Delta=14.88$~meV because all
changes in magnetic field are compensated by a change
in temperature as required by the modified Brillouin function [Eq.~(\ref{eq_delta})].
For this reason, the splitting of the peaks are identical between adjacent curves.
However, since each curve is acquired at a different $B$, the Zeeman splitting
in the injector $h$, which is temperature independent, changes.
As a result, the relative amplitude of the two peaks changes.
At low temperatures, the applied field is small
and the injected current is spin unpolarized.
When $B$ is increased (and $k_B T$ in order to keep $\Delta$ constant), 
the injector becomes polarized
and the majority spin current peak increases whereas
at the same time the minority peak decreases.
In the inset of Fig.~\ref{figmqw_th}
we show the expected polarization detected by the RTD, which
exhibits a roughly linear dependence with $p$, thus
reinforcing the idea that the polarization
can be measured via the change of the peak height.

We emphasize that the operating principle of this device lies in
the giant Zeeman splitting in the well which allows us to separate
the current into spin-up and spin-down flows and that the increase
of $B$ only serves as a tool to generate spin polarizations in the
injector. Obviously, any other polarization generating method would play a
similar role.

\section{Conclusion}
We have addressed nonlinear electron transport in
magnetically doped II--VI resonant tunneling diodes.
We have presented experimental data which shows that such
devices can work as spin filters when the quantum well is magnetic.
We have presented a simple theoretical model which accounts for
the observed behavior and have proposed two devices in which
spin effects are important.

The first device consists of a RTD with an emitter doped with 
Mn and a normal quantum well. Our model shows an enhancement
and a voltage shift of the resonance current peak when the applied
magnetic field increases. The results are consistent with a giant
Zeeman splitting in the injector since the current saturates
at a few Tesla and the temperature dependence follows
the magnetization of a paramagnetic system.
The Zeeman splitting induced redistribution of spin carriers in the injector
leads to a modification of the conduction band structure of the device which
is responsible for the changes in the transport properties.
The results we obtain are reminiscent of Ref.~\cite{cho92}, where
the In content of a GaInAs emitter in a III-V RTD was varied.
As a consequence, the band alignment changes and the peak current
increases. In our case, the increase is
due exclusively to the giant spin splitting in the injector,
offering the unique possibility of producing high peak
currents which arise from spin effects only.
Thus, the current increase can be tuned with a magnetic field
without changing the sample parameters.

The second device is a RTD where the central well is magnetically doped
and the applied fields are high. Then, we have presented evidence
that such a RTD can work as a
detector of spin polarized currents. The Mn impurities in the well
give rise to a giant Zeeman splitting which manifests itself in
a splitting of the $I$--$V$ characteristic. 
Further increase of the magnetic field leads to a change in the amplitude
of the current peaks which is opposite for each spin index.
Thus, this amplitude change is employed to estimate the injector's spin polarization.

The theoretical model has considerable latitude for including complicated
effects. It has been shown in Ref. \cite{san02} that electron-electron
interactions (even at the mean-field level) can produce multistability
of states with distinct spin polarizations for a given bias voltage.
Depending on the carrier density, in multiple quantum well systems
there may arise self-sustained oscillations of the spin polarization
which occur without any time-dependent forcing \cite{bej03}.
In addition, a desirable property of any transport theory dealing with
spin-polarized currents is that spin relaxation processes should be
taken into account. These mechanisms can be rather slow in II--VI
materials but their role can be discussed within a phenomenological
model \cite{san02}.

In closing, we believe that our discussion opens new research avenues
toward an all-electrical generation, manipulation and detection
of spin polarized currents in nanodevices.

\section*{Acknowledgment}
We thank A. Slobodskyy, T. Slobodskyy and D.L. Supp
for collaboration in the experiments and V. Hock for sample fabrication.

\begin{biography}{David S\'anchez}
received his PhD degree in physics from the Autonomous University of Madrid in 2002 for theoretical work in nonlinear and spin-related properties of electronic nanodevices. He is currently an assistant researcher with the Department of Physics at the University of the Balearic Islands. His main interests include nonlinear mesoscopic transport, strongly correlated nanostructures and spintronics.
\end{biography}

\begin{biography}{Charles Gould}
completed his PhD in mesoscopic transport physics in a co-operative program between Sherbrooke University and the National Research Council of Canada (CNRC) in 2000. Since September 2000, he has been working as a post-doc/scientific assistant in the group of Prof. Molenkamp, doing transport work in semiconductor spintronics with a large focus on (Ga,Mn)As-based devices.
\end{biography}

\begin{biography}{Georg Schmidt}
did his PhD in Aachen working on Si/SiGe deposition and nanolithography. In 1996 he joined the group of Prof. Molenkamp in Aachen and started the activities on spintronics and nanolithography. Since 1999 he is in Wuerzburg at the Chair of Prof. Molenkamp, where he is the head of the workgroup for spintronics and nanolithography. His main focus is on spin transport and ultra high resolution lithography.
\end{biography}

\begin{biography}{Laurens Molenkamp's}
background includes the optical and quantum transport studies in semiconductor nanostructures and spintronics. He was appointed Professor of Experimental Physics (C3) at the RWTH in Aachen, Germany, in 1994. He moved to Wuerzburg University in April 1999 as the Chair of Experimental Physics (EP3), where he continues research in spintronics and spin-related phenomena.
\end{biography}






\end{document}